# Researches on a reactor core in heavy ion inertial fusion

S. Kondo[1*], T. Karino[1], T. Iinuma[1], K. Kubo[1], H. Kato[1],

S. Kawata[1*], A. I. Ogoyski[2]

[1]*Graduate School of Engineering, Utsunomiya University, Yohtoh 7-1-2, Utsunomiya 321-8585, Japan*

[2] *Department of Physics, Technical University of Varna, Ulitska, Studentska 1, Varna, Bulgaria*

Abstract

In this paper a study on a fusion reactor core is presented in heavy ion inertial fusion (HIF), including the heavy ion beam (HIB) transport in a fusion reactor, a HIB interaction with a background gas, reactor cavity gas dynamics, the reactor gas backflow to the beam lines, and a HIB fusion reactor design. The HIB has remarkable preferable features to release the fusion energy in inertial fusion: in particle accelerators HIBs are generated with a high driver efficiency of ~30-40%, and the HIB ions deposit their energy inside of materials. Therefore, a requirement for the fusion target energy gain is relatively low, that would be ~50 to operate a HIF fusion reactor with a standard energy output of 1GW of electricity. In a fusion reactor the HIB charge neutralization is needed for a ballistic HIB transport. Multiple mechanical shutters would be installed at each HIB port at the reactor wall to stop the blast waves and the chamber gas backflow, so that the accelerator final elements would be protected from the reactor gas contaminant. The essential fusion reactor components are discussed in this paper.

Introduction

In inertial confinement fusion (ICF), a driver efficiency and its repetitive operation with several Hz ~ 20 Hz or so are essentially important to constitute an ICF reactor system. Heavy ion beam (HIB) driver accelerators have a high driver energy efficiency of 30-40 % from the electricity to the HIB energy. In addition, high-energy accelerators have been operated repetitively daily. The high driver efficiency relaxes the requirement for the fuel target gain. In HIF the target gain of ~50 would be needed to construct HIF fusion reactor systems.

In ICF target implosion, the requirement for the implosion uniformity is very stringent, and the implosion non-uniformity must be less than a few % (Emery et al., 1982; Kawata et al., 1984). Therefore, it is essentially important to improve the fuel target implosion uniformity. In general, the target implosion non-uniformity is introduced by a driver beams' illumination non uniformity, an imperfect target sphericity, a non-uniform target density, a target alignment error in a fusion reactor, et al. The target implosion should be robust against the implosion non-uniformities for the stable reactor operation.

The HIB must be transported in a fusion reactor, which would be filled by a debris gas plasma. The reactor radius would be 3~5 m or so. From the beam exit of the accelerator the HIB should be transported stably. The HIB ion is rather heavy. For example, $Pb^+$ ion beams could be a promising candidate for the HIF driver beam. Fortunately the heavy ions are transported almost ballistically with straight trajectories in a long distance. Between the HIB ions and the background electrons, the two-stream and filamentation instabilities may appear, and simple analyses confirm that the HIBs are almost safe from the instabilities' influences. However, the HIB's self-charge may contribute to a slight radial expansion of the HIBs especially near the fuel target area due to the neutralized electrons' heating by the HIB radial compression during the HIB's propagation in a reactor (Kawata et al., 2005). So the HIB charge neutralization is also discussed in the paper.

The HIB uniform illumination is also studied, and the target implosion uniformity requirement requests the minimum HIB number: details HIBs energy deposition on a direct-drive DT fuel target shows that the minimum HIBs number would be the 32 beams (Miyazawa et al., 2005). The 3-dimensional detail HIBs illumination on a HIF DT target is computed by a computer code of OK (Ogoyski et al, 2004; Ogoyski et al., 2004; Ogoyski et al., 2010). The HIBs illumination non-uniformity is also studied in detail. One of the study results shows that a target misalignment of ~100μm is tolerable in fusion reactor to release the HIF energy stably. In addition, a preliminary HIF reactor component study is also presented.

Issues in heavy ion inertial fusion about target and beam pluse

In this section key issues in HIF are summarized first. The fuel target design should be conducted further toward a robust fuel implosion, ignition and burning. The HIF target design is quite different from the laser fusion target due to the relatively long range (the order of ~1mm) of the energy deposition (Kawata et al., 1984; Kawata et al.).

An example direct-drive fuel target is presented in Fig. 1. The target should be compressed to about one thousand of the solid density to reduce the driver energy and to enhance the fusion reactions (Atzeni et al., 2009). The target should be robust against the small non-uniformities caused by the driver beams' illumination non-uniformity, a fuel target alignment error in a fusion reactor, the target fabrication defect, et al. The ICF reactor operation frequency would be 10~15Hz or so. So the stable target performance should be realized. The HIB stopping range is rather long, and the HIB beam energy is deposited mainly at the end area of the beam ion stopping range due to the Bragg peak effect, which is originated from the nature of the Coulomb collision. The interaction of the HIB ions could be utilized to enhance the HIB preferable characteristics. The HIB ion interaction is relatively simple, and is almost the classical Coulomb collision, except the plasma range-shortening effect (Ichimaru, 2004).

However, the HIBs illumination scheme should be studied intensively to realize a uniform energy deposition in a HIF target. The ICF target implosion uniformity must be less than a few % (Emery et al., 1982; Kawata et al., 1984). The uniformity requirement must be fulfilled to release the fusion energy. The multiple HIBs should illuminate the HIF target with a highly uniform scheme during the imploding DT shell acceleration phase.

In addition, the HIB pulse shape should be also designed to obtain a high implosion efficiency $\eta$. An example $Pb^+$ HIBs pulse shape is presented in Fig. 2. The pulse shape consists of a low-intensity foot pulse, a ramping part to the peak intensity and the high-intensity main pulse. The foot pulse generates a weak shock wave in the target material and the DT fuel, and the first shock wave kicks the low temperature DT liquid fuel inward. When no foot pulse is used, the main pulse with the high intensity generates a strong first shock wave inside of the DT fuel and increases the DT adiabat. The DT fuel preheat would be induced. The foot pulse length and the ramping time are designed to reduce the entropy increase in the DT fuel layer. The first weak shock wave is not caught by the second and third stronger shocks inside of the DT fuel layer. At the inner edge of the DT fuel layer the shocks should be overlapped so that the efficient fuel acceleration and compression are accomplished during the implosion. The detail pulse shape should be designed for each target design.

The reactor design is also another key issue in ICF (Emery et al., 1982; Kawata et al.,1984; Kawata et al., 2005) . The first wall could be a wet wall with a molten salt or so or a dry wall. The reactor design must accommodate a large number of HIBs beam port, for example, 32 beam ports. At the first wall and the outer reactor vessel the beam port holes should have mechanical shutters or so to prevent the fusion debris exhaust gas toward the accelerator upstream. In addition, the target debris remains inside of the first wall or mixes with the liquid molten salt, which may be circulated. The target debris treatment should be also studied as a part of the reactor design.

The HIB accelerators have the high controllability and flexibility for the particle energy, the beam pulse shape, the pulse length, the ion species, the beam current, the beam radius, the beam focusing and the beam axis motion. However, the high-current operation may need additional studies for the fusion reactor design. The a few~10kA HIB may induce an additional HIB divergence, a beam loss and an electron cloud generation in the accelerator. The high-current and high-charge HIB generation and transport should be studied carefully to avoid the uncertainty in the HIB accelerator.

## Target gain requirement for HIF reactor system

A target energy gain required for an energy production is evaluated by a reactor energy balance in ICF shown in Fig. 3. The driver pulses deliver an energy $E_d$ to a target, which releases fusion energy $E_{\text{fusion}}$. The energy gain is $G = E_{\text{fusion}}/E_d$. The fusion energy is first converted to electricity by a standard thermal cycle with an efficiency of $\eta_{\text{th}}$. A fraction f of the electric power is circulated to the reactor system operation and the driver system, which converts it to the HIB energy with an efficiency of $\eta_d$. The energy balance for this cycle is written by $f\eta_{th}\eta_d G > 1$. Taking $\eta_{th} = 40\ \%$ and requiring that the circulated-energy fraction f of electrical energy should be less than 1/4, we find the condition $G\eta_d > 10$. For a driver efficiency in the range of $\eta_d = 30 \sim 40\%$, the condition G > ~30 is required for power production. Therefore, the preferable fusion target gain would be G ~ 50~70 in HIF. When the HIF reactor system operation is about 10~15 Hz, a 1GW HIF power plant can be designed.

## HIB stable transport in a fusion reactor

The HIB should be focused and transported in a fusion reactor against the beam space charge onto a fuel pellet at the reactor center. The target radius is the order of mm. One of the promising transport schemes is the neutralized ballistic transport, in which preformed-plasma electrons or wall-emitted electrons neutralize the HIB space charge. On the other hand, the HIB ion number density increases from $n_{b0} \sim 10^{11}$–$10^{12}$cm$^{-3}$ at a HIB port entrance to $100 \sim 200 \times n_{b0}$ at the fuel pellet position. As a practical HIB neutralization method in an HIF reactor, an insulator annular tube guide was proposed at the final transport part, through which a HIB is transported (see Fig. 4) (Kawata et al.,2003). A local electric field created by the intense HIB induces local discharges, and a plasma is produced at the annular insulator inner surface. The electrons are extracted from the plasma by the HIB net charge. The electrons neutralize the HIB charge, and move together with the HIB in the reactor chamber (see Fig. 5-8).

In addition, the chamber background electron density should be always larger than or equal to the HIB number density: $n_{ce} > Z_b n_b$. Here, $n_{ce}$ is background electrons density, $Z_b$ is degree of ionization and $n_b$ is the HIB density. In our design, the fusion reactor is filled with helium gas, and its density is $\sim 10^{16} \sim 10^{17} [1/cm^3]$. In this case, the HIB space charge is always neutralized in the reactor well. The instability analyses were also performed including the two-stream instability and the filamentation instability. The maximum growth rate of the two-stream instability between the HIB ions and the background electrons is given by Eq. (1) (Okada et al., 1981).

$$\gamma_{max} = -\nu + \sqrt{\frac{\pi}{2}} \frac{\omega_b^2}{\omega_p} \frac{\left(V_b - u_b/\sqrt{2}\right)^2}{u_b^2} \exp\left(-\frac{1}{2}\right) \qquad (1)$$

Here the collision frequency is evaluated by the Coulomb interaction between the background electron and the HIB ions, whose speed is $V_b$ and thermal speed is $u_b$. The HIB ion plasma frequency is denoted by $\omega_b$, and $\omega_e$ is the background electron plasma frequency. The maximum growth rate of the filamentation instability is given by Eq. (2) (Okada et al., 1981).

$$\gamma_{max} = 2 \frac{\omega_b^2}{\omega_p^2} \frac{V_b^2}{u_b^2} \nu \qquad (2)$$

The two-stream and filamentation instability analyses are summarized in Fig. 9. It was found the two-stream instability is stabilized, and the filamentation instability could be unstable. However, the product of the growth rate and the HIB transport time is small, that is, $\gamma\tau \leq 5$ or so.

Reactor cavity gas

In this section a reactor chamber gas dynamics is studied by using a simple analysis (Bondorf et al., 1978; Zel'dovich et al., 2002). A simple spherically symmetric fireball in a fusion reactor is estimated by the equation of continuity, the equation of motion and the equation of state for the adiabatic expansion:

$$\frac{\partial \rho}{\partial t} + \left(\frac{1}{r^2}\right) \cdot \frac{\partial (r^2 \rho u)}{\partial r} = 0, \qquad (3)$$

$$\frac{\partial u}{\partial t} + u \frac{\partial u}{\partial r} = -\frac{1}{\rho} \cdot \frac{\partial p}{\partial r}, \qquad (4)$$

$$\frac{p}{\rho^r} = \text{constant} \qquad (5)$$

One of the solutions is the ZBG (Zimanyi, Bondorf and Garpman) solution (Bondorf et al., 1978):

$$R^2(t) = R_0^2 + <u_t>^2 (t - t_0)^2, \qquad (6)$$

$$\rho(t,r) = \rho_0 \left(\frac{R_0^3}{R^3}\right) \cdot \left(1 - \frac{r^2}{R^2}\right)^\alpha, \qquad (7)$$

$$T(t,r) = T(t_0,r) \cdot \left(\frac{R_0^2}{R^2}\right), \qquad (8)$$

$$\alpha = \frac{m<u_t>^2}{2T_0} - 1 \qquad (9)$$

In this work $\alpha=1$ is employed, and then we obtain a simple adiabatic expansion of the blast wave gas. In one shot of the DT fuel target implosion, ignition and burning, the initial temperature would be ~10keV or so. In this case the sound speed of $C_s \sim <u_t> \sim 10^8$ cm/s or so. Therefore, the blast wave traveling time $\tau$ in a fusion reactor chamber may be $\sim 500[cm]/10^8[cm/s] \sim 5[\mu sec]$. The target-debris chamber gas density would be the order of 1~a few torr or so, which corresponds to the gas number density of $8 \times 10^{14}[1/cm^3]$, after the blast wave reaches the chamber wall. Without the chamber gas supply, the expansion of the chamber gas after the target burning may make the target-debris gas dilute. During each shot, the chamber gas may be actively supplied to keep the high-density chamber gas in the chamber. When the reactor system operated with a 10~15 Hz, the blast wave propagation is too fast to give a significant interaction between the blast waves. The blast wave expands vary fast in ~5μs, and the time interval for two DT fuel target implosions is about 1/15~1/10 s.

The gas density should be high enough to compensate the focusing HIB charge. This is a good way for the HIB transport in a fusion reactor chamber. However, for accelerators and for the final focusing elements in the accelerator final sections, a part of the exhaust gas and debris coming up may have influences. The vacuum of the accelerator part should be kept to the low pressure to avoid the HIB ions scattering and the halo formation.

In the HIF reactor system we may have the ceramics HIB transport annular guide (see Figs. 5-8), whose inner surface would absorb the upcoming exhaust chamber gas to the accelerator final section. The ceramics material surface has many small holes, which absorb the debris gas and vapor (Nishiyama et al., 1978; Hanamori et al., 1998; Kawata et al., 2003; Kato et al., 1995). If the beam ports at the chamber wall have no mechanical shutters to stop the blast waves and the chamber gas flow, the accelerator final elements may meet the contamination. So at the final part of each accelerator near the chamber wall, each accelerator final part may several mechanical shutters, and two pairs of the mechanical shutters would confine the exhaust gas and absorb it at the ceramics guide inner surface. The absorbed gas is reused to produce a plasma at the ceramics annular guide inner surface to supply the electrons to neutralize the next HIB space charge (see Fig. 5) (Nishiyama et al., 1978; Hanamori et al., 1998; Kawata et al., 2003; Kato et al., 1995). This is a realistic solution to keep the accelerator final section clean from the chamber gas and debris. Figure 10 show an example concept for the accelerator protection system.

In this example in Fig. 10, the front shutter rotates with the speed of 100 rotations per second. The first shutter stops the most part of the reactor gas backflow to the accelerator final part. The second shutter rotates with 10 rotations per second. The reactor gas speed is about $C_s \sim 6.0 \times 10^5 [cm/s]$ at the reactor wall, where the beam ports are located. Some part of the reactor exhaust gas is confined at the beam port section isolated by the two shutters. When the annular ceramics insulator guides are installed at the final beam port section, the reactor exhaust gas would be absorbed perfectly and the accelerator vacuum is kept well. In order to absorb the reactor backflow gas at the accelerator final part, several pairs of the mechanical shutters shown in Fig. 10 would be required to keep the accelerator vacuum.

Figure 11 shows our present HIF reactor design, which has a molten-salt liquid first wall. The helium chamber gas is also injected to keep the HIB charge neutrality. In this design 32 HIBs are employed, and the HIB ports are also displayed in Fig. 11. At each port the mechanical shutter pairs are installed to protect the accelerator final section from the reactor exhaust gas back flow. The operation frequency would be 12 Hz and the fusion thermal energy output is 3GW to produce 1GW electricity. Our further reactor conceptual studies in HIF will be performed in the near future to confirm the HIF reactor system feasibility.

## Conclusions

In this paper we presented our recent research results relating to the HIF reactor core, including the fuel target implosion uniformity requirement, the target gain requirement, the HIB space charge neutralization method, the HIB focusing feature, the HIB instability study in a fusion reactor chamber, and the preliminary HIB reactor design. So far our studies have presented that the HIF reactor concept has no critical defect in order to construct a stable and reliable fusion power reactor system. We found also that the HIB reactor system is relatively simple in all aspects required in the HIF system. This finding would be preferable for our future HIF reactors.

Acknowledgements

The work was partly supported by JSPS, MEXT, CORE (Center for Optical Research and Education, Utsunomiya University), ILE / Osaka University, and CDI (Creative Department for Innovation, Utsunomiya University). The authors also would like to extend their acknowledgements to friends in HIF research group in Japan, in Tokyo Inst. of Tech., Nagaoka Univ. of Tech., KEK and also in HIF-VNL, U.S.A.

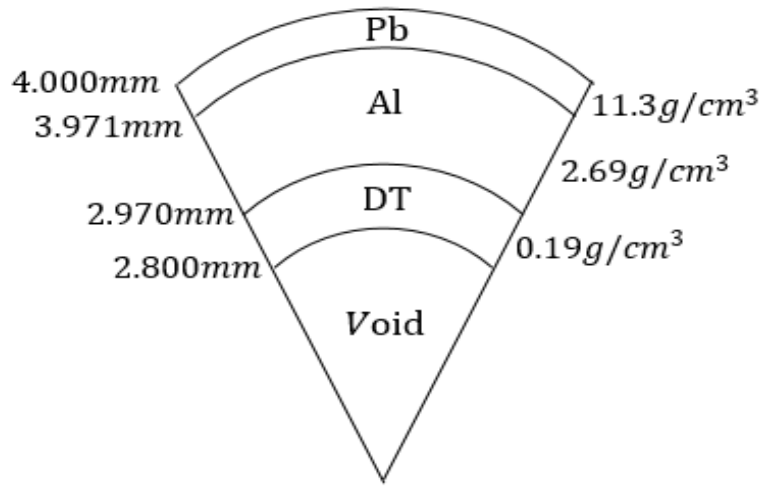

Fig.1 An example fuel target structure in heavy ion inertial fusion.

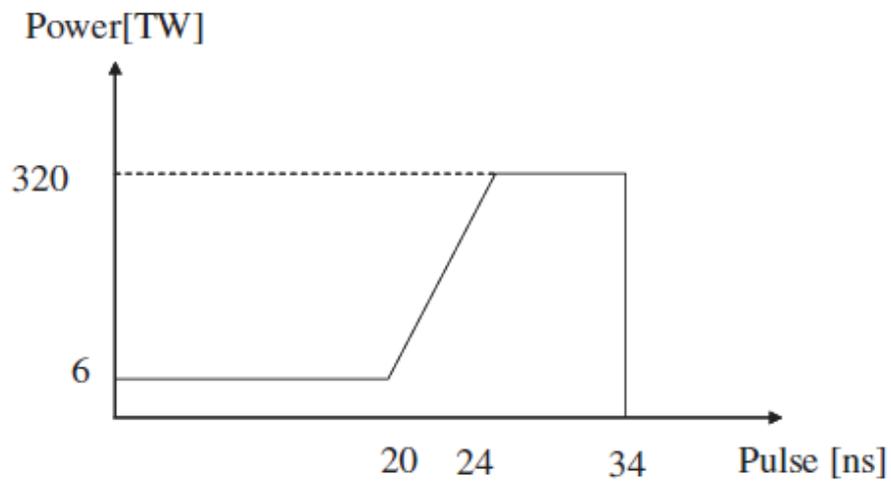

Fig.2 An input heavy ion beam pulse. The HIB pulse consists of the low power part (foot pulse) and the high power one (main pulse).

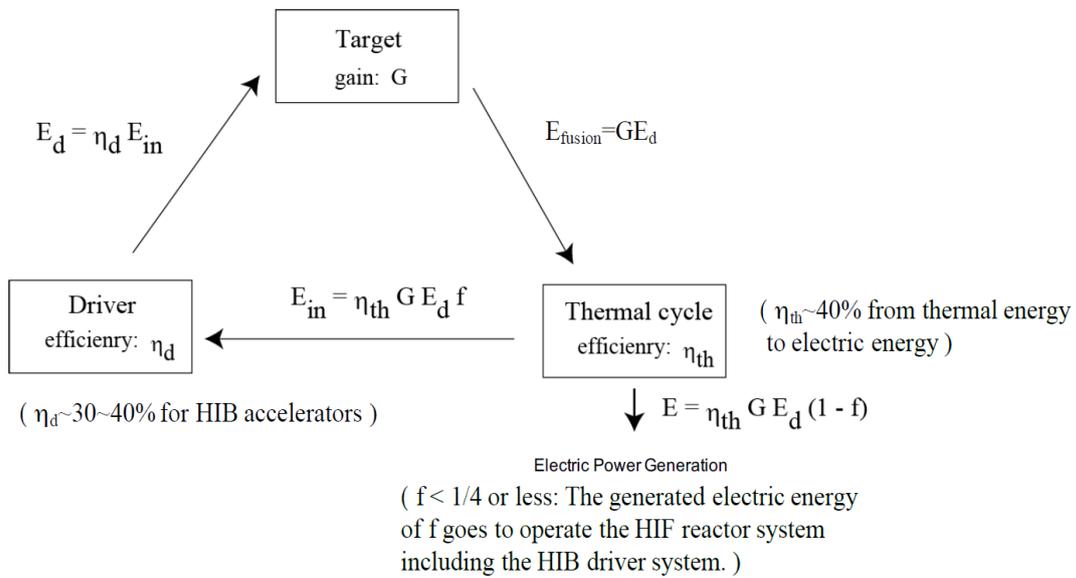

Fig.3 Energy balance in ICF reactors.

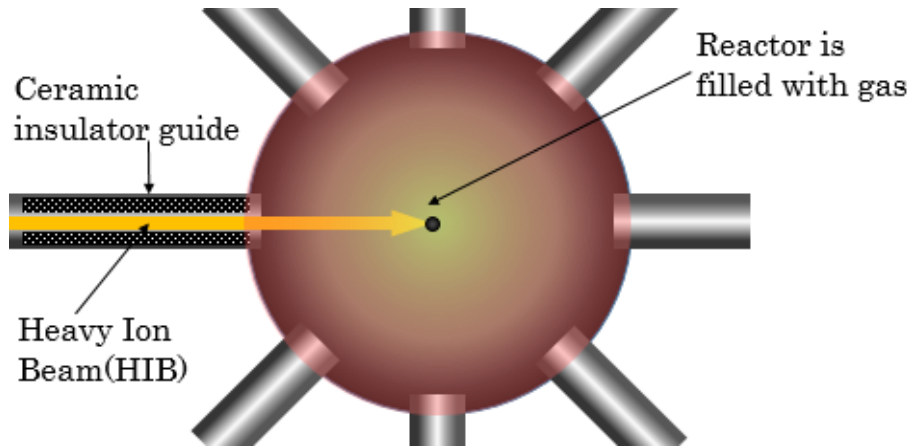

Fig.4 A conceptual diagram of a HIF reactor. First, beam is transported through the ceramics Insulator guide. Second, beam transport through the reactor gas. Third, after the fusion reactions, the reactor gas expands in a reactor. Finally, the generated gas will enter the beam port.

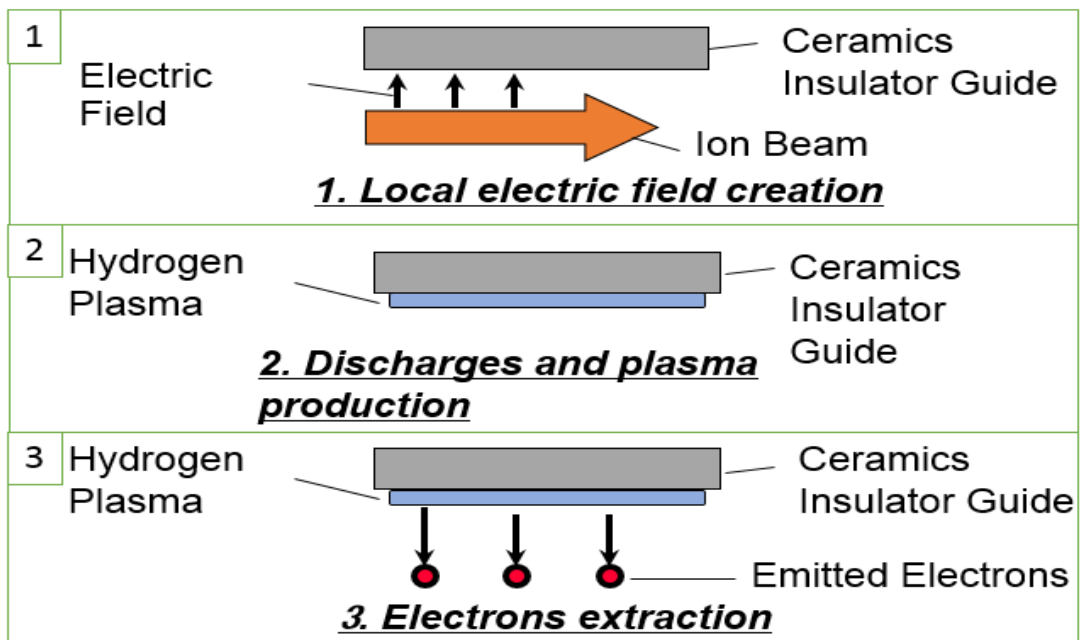

Fig.5 Electron extraction mechanism to compensate the HIB space charge. An intense HIB creates a strong electric field at the insulator ceramics annular guide inner surface, and induces the electric discharges, which produces a plasma at the annular insulator surface. From the created plasma the HIB extract electrons to neutralize the HIB net charge in the regulated way.

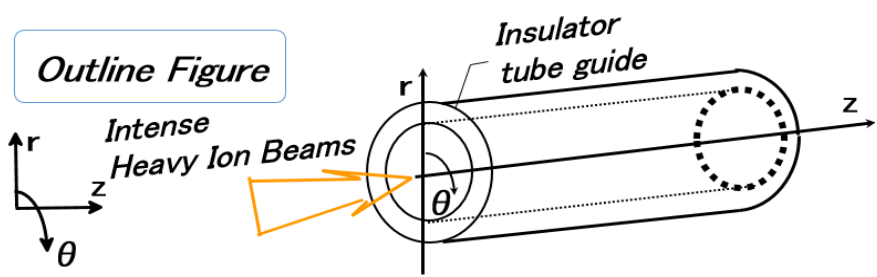

Fig.6 Concept of an annular insulator guide for HIB charge

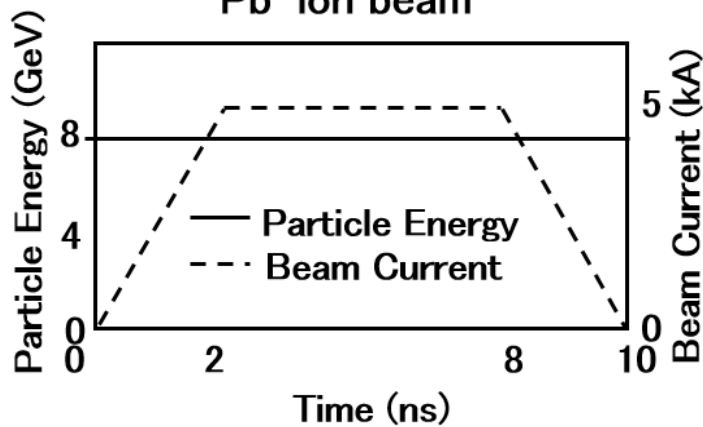

Fig.7 Input $Pb^+$ ion beam waveform. The $Pb^+$ ion-beam parameter values are as follows: the maximum current is 5 kA, the particle energy is 8 GeV, the pulse width is 10 ns, and the rise and fall times are 2.0 ns.

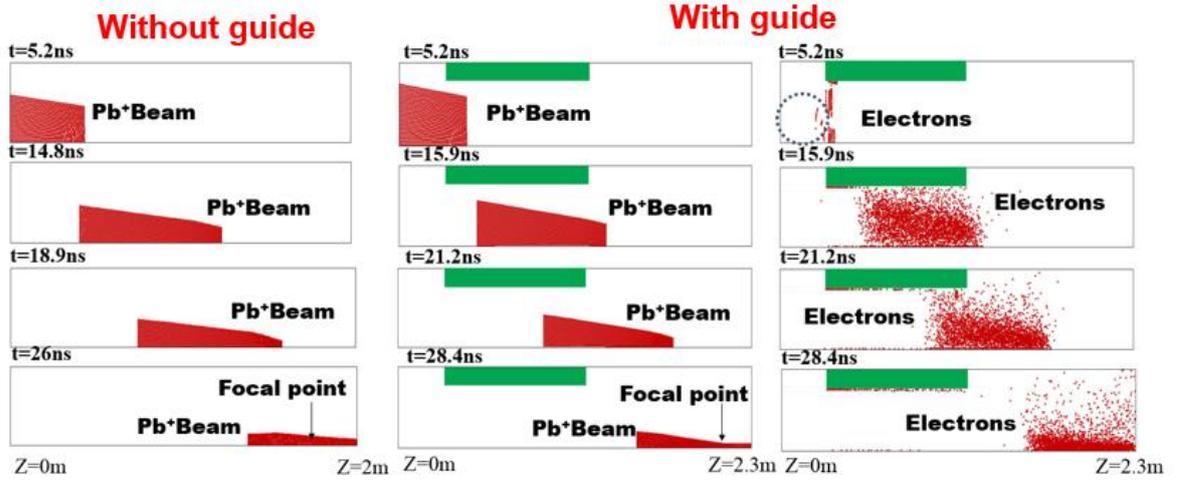

Fig.8 $Pb^+$ ion particle map and electron map in the case without the annular insulator guide and with the annular insulator guide. The focal radius is 2.4 mm at Z=2.1 m with the guide, and 6.0 mm at Z=2.1m without the guide.

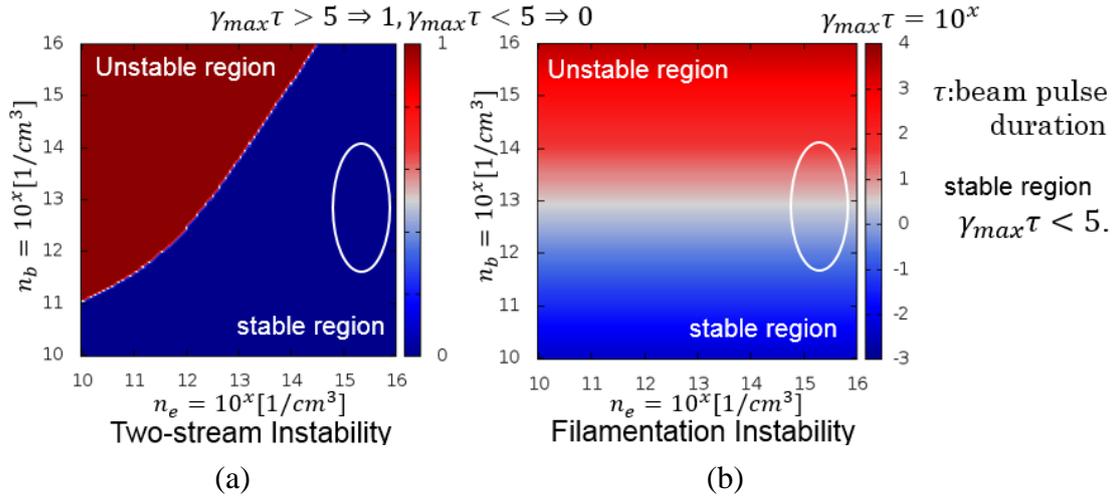

Fig.9 (a) shows the growth rate of the two-stream instability and (b) shows the growth rate of the filamentation instability. The two-stream instability is always stable in the reactor. However, the product of the growth rate and the HIB transport time is small, that is, $\gamma\tau \leq 5$ or so

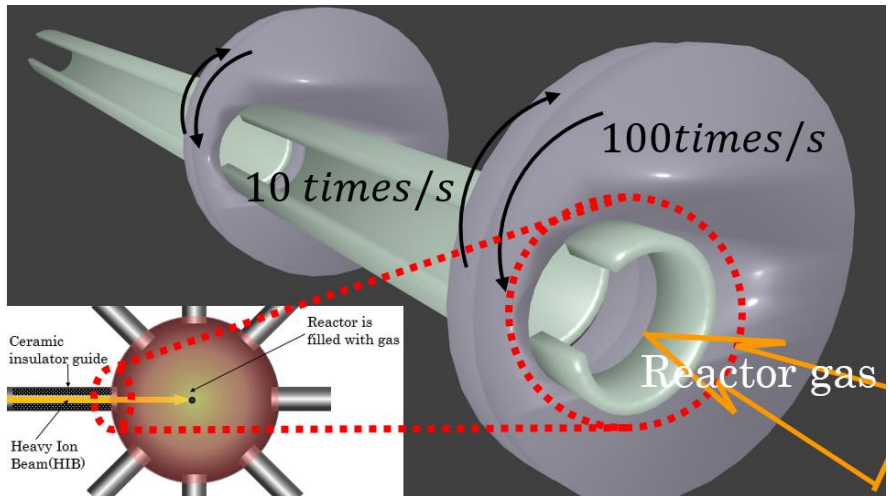

Fig. 10 Mechanical shutters to protect the accelerator vacuum from the reactor exhaust gas backflow.

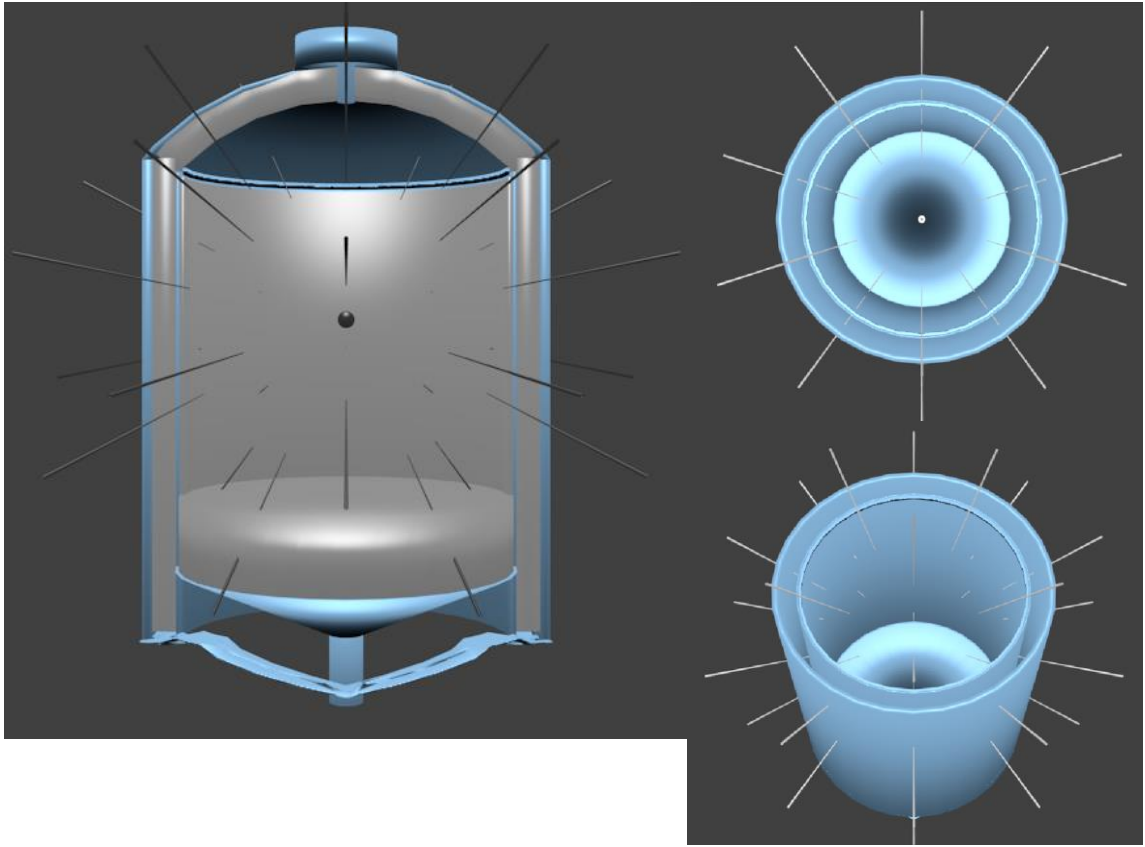

Fig. 11 A concept of a HIF fusion reactor, which has a molten-salt liquid first wall. The helium chamber gas is also injected to keep the HIB charge neutrality. In this design 32 HIBs are employed, and the HIB ports are also displayed in the figure. At each port the mechanical shutter pairs are installed to protect the accelerator final section from the reactor exhaust gas backflow.